\documentclass[aps,prd,nofootinbib,preprintnumbers,floatfix,twocolumn,superscriptaddress]{revtex4}

\usepackage{amsmath,amssymb,amsfonts,bm}
\usepackage{empheq}
\usepackage{physics}

\usepackage[svgnames,dvipsnames]{xcolor}
\usepackage{graphicx}
\definecolor{NewBlue}{rgb}{0.1, 0.1, 0.7}
\definecolor{NewRed}{rgb}{0.7, 0.1, 0.1}
\usepackage[colorlinks,
    linkcolor=Maroon,
    citecolor=NewBlue,
    urlcolor=NewRed]{hyperref}

\usepackage{cleveref}

\newcommand{\beq}{\begin{equation}} 
\newcommand{\eeq}{\end{equation}} 
\newcommand{\bea}{\begin{eqnarray}} 
\newcommand{\eea}{\end{eqnarray}} 
\def\benu{\begin{enumerate}}
\def\eenu{\end{enumerate}}
\def\nn{\nonumber}

\def\pa{{\partial}}
\def\l{\left}
\def\r{\right}
\def\d{{\rm d}}
\def\f{\frac}

\def\k{{\bf k}}
\def\x{{\bf x}}
\def\km{{\rm k}}

\def\e{{\rm e}}

\begin{document}
 
\title{Lorenz gauge and the Poisson bracket in canonical electromagnetism}
\author{D. Jaffino Stargen}
\email{jaffinostargend@gmail.com}
\affiliation{Department of Mechanical Engineering, 
Massachusetts Institute of Technology, Cambridge, MA 02139, USA}
\begin{abstract}
In treatments of electromagnetism, it is often tacitly assumed that the vector
potentials of the field and their conjugate momenta satisfy the canonical Poisson
bracket relations, despite the fact that the components of the vector potential
are constrained by gauge conditions. Here I explicate how this comes about by
imposing Poisson bracket relations on the independent field variables remaining
after the Lorenz gauge constraint is accounted for. The naively assumed Poisson
brackets happens to be correct even \emph{after} gauge fixing, owing to a
conspiracy between the gauge conditions and the principle of relativity.
\end{abstract}
\maketitle

\section{Introduction}

In the Hamiltonian view of classical systems
\cite{Whitt1917,Sommer1964,LanLif1976,Arnold1989}, the state of a system is 
specified by a set of $n$ generalized coordinates $\vb{q}=\{q_i\}$ and their
conjugate momenta $\vb{p}=\{p_i\}$, assumed to be unconstrained and satisfying
the Poisson bracket relation $\{q_i, p_j\} = \delta_{ij}$; time evolution of
the state is described by Hamilton's equations: $\dot{q}_i = \{q_i, H\}$ and
$\dot{p}_i = \{p_i, H\}$. Thus, in this scheme, the Poisson bracket between
the $n$ \emph{independent} pairs of variables (``degrees of freedom'') plays
a crucial role.

If the degrees of freedom are not independent, i.e. if they are constrained,
$\phi(\vb{q},\vb{p})=0$, then the Hamiltonian formalism has to be
amended \cite{Dirac58,Dirac-1964,Teitelboim:1976,Sund1982,Sudarsan-Mukunda:2015}: 
either the Poisson bracket has to be modified, or, if they are retained by
treating the variables as independent, then the constraints $\phi=0$ need to be
imposed as auxiliary conditions.

The classical theory of the electromagnetic field naturally presents a constrained
problem. If the system is to be treated in an explicitly Lorentz covariant manner,
then the appropriate field coordinates are the vector potentials $\{A^\mu\}$.
However, not all configurations of the potentials are physically distinct,
i.e. a gauge constraint of the form $\phi(A^\mu)=0$ appears between them.
Paralleling the discussion above, one of two possibilities can address these
constraints. The Poisson bracket suitable, had the field variables been independent,
\begin{equation}\label{PB1}
  \{A^{\mu}(t,\x),\pi^{\nu}(t,\x')\}=\eta^{\mu\nu} \delta(\x-\x'),
\end{equation}
needs to be modified to account for the gauge constraints, where the signature
of the Minkowski
metric is $\eta=(-1,1,1,1)$ \cite{Sudarsan-Mukunda:2015,Greiner,
Itzykson-Zuber:2006,Prokhorov:1988}.
Or, if they are to be retained as above, the gauge constraint needs to be imposed
separately, and further, it must be ensured that once imposed, they remain valid 
for all times, i.e. $\{\phi(A^\mu),H\}=0$. Both these routes can be explored by using
Dirac's theory of constrained Hamiltonian mechanics 
\cite{Teitelboim:1976,Sund1982,Scherer:1987,GitTyut,Gieres:2021}.

A less sophisticated, and arguably more transparent, approach is conceivable:
eliminate the gauge constraint explicitly in order to identify the true
independent degrees of freedom of the field, and impose Poisson brackets on those.
Although not apriori obvious, such a procedure, if performed consistently, agrees
with \cref{PB1}. The primary purpose of our exposition is precisely to demonstrate
how this comes about, i.e., how the Poisson bracket structure of electromagnetism
[\cref{PB1}] is upheld \emph{after} the gauge constraint is accounted for. 
As it happens, this relies on the principle of relativity
in an essential manner.

The exposition is organized as follows: in \Cref{sec:Setup} we describe the
appropriate degrees of freedom corresponding to the electromagnetic field system;
\Cref{sec:PB} explicates how the Poisson bracket corresponding to the field
variables, $A^{\mu}$ and $\pi^{\mu}$, is modified due to the gauge constraint,
and indeed how the Poisson bracket preserves its structure even after the gauge
constraint is taken into consideration; and we conclude in \Cref{sec:Conclusion}.
\section{Canonical electromagnetism}\label{sec:Setup}
We start with the Lagrangian density of the electromagnetic field as
\cite{cohen,Greiner}
\begin{equation}\label{Lagrangian}
 {\cal L}=-\f{1}{2} \pa_{\mu} A_{\nu} \pa^{\mu} A^{\nu},
\end{equation}
where the four-potential, $A^\mu$, are taken to be the (generalized) coordinates,
whose canonically conjugate momentum, $\pi_{\mu}$, is \begin{equation}\label{eqn:pi}
 \pi_{\mu} \equiv \f{\pa {\cal L}}{\pa (\pa_{0} A^{\mu})}=\pa_{0} A_{\mu}.
\end{equation}
However, due to the gauge constraints involved, the field coordinates, $A^\mu$,
are not independent of each other. The gauge constraint of our choice is of the
Lorenz form, $i.e.$,
\begin{equation}\label{Lorenz}
  \partial_\mu A^\mu = 0.
\end{equation}

In order to render the subsequent discussion as simple as possible, 
we decompose the field coordinate, $A^{\mu}$, as
\begin{equation}\label{A_mu}
 A^{\mu}(\vb{x},t)=\sum_{\lambda=0}^{3} \int \d^3\k
 \varepsilon^{\mu}_{(\lambda)} q^{(\lambda)}_{\k}(t) e^{-{\mathrm i}\k\cdot \x}, 
\end{equation}
where $\varepsilon^{\mu}_{(\lambda)}$ are polarization vectors, with $\mu$ and
$\lambda$ signifying the spacetime and polarization indices, respectively;
the variables, $q^{(\lambda)}_{\k}$, capture the dynamical degrees of freedom
corresponding to the field, and are complex variables in general.
Since the vector potential, $A^{\mu}$, is real, the dynamical variables,
$q^{(\lambda)}_{\k}$,
and its complex conjugate, $q^{*(\lambda)}_{\k}$, are related by
$q^{*(\lambda)}_{\k}=q^{(\lambda)}_{-\k}$.

Therefore, the field Lagrangian, $L$, in terms of the dynamical
variables, $q^{(\lambda)}_{\k}$, can be obtained from the Lagrangian density,
${\cal L}$, of the field [\cref{Lagrangian}] as \begin{equation}\label{eqn:Lagrangian}
 L \equiv \int \d^3\x~{\cal L}(t,\x) = \int \d^3\k~L_{\k},
\end{equation}
where
\begin{equation}
 L_{\k} \equiv \eta_{\lambda\lambda'}
 \biggl({\dot q}_{\k}^{(\lambda)} {\dot q}_{-\k}^{(\lambda')}
 -|\k|^2 q_{\k}^{(\lambda)} q_{-\k}^{(\lambda')}\biggr).
\end{equation}
Since the dynamical variables, $q^{(\lambda)}_{\k}$, are independent of each
other prior to taking gauge constraint into consideration, the Lagrangian,
$L_{\k}$, corresponding to a field mode, labelled by $\k$, is composed of four
{\it independent} dynamical systems corresponding to the dynamical variables,
$q^{(\lambda)}_{\k}$. The momentum variables canonically conjugate to the
variables, $q^{(\lambda)}_{\k}$, can be found to be
\begin{equation}
 p_{\k(\lambda)} \equiv \f{\pa L_{\k}}{\pa {\dot q}^{(\lambda)}_{\k}}
 ={\dot q}_{-\k(\lambda)},
\end{equation}
\begin{equation}
 p_{-\k(\lambda)} \equiv \f{\pa L_{\k}}{\pa {\dot q}^{(\lambda)}_{-\k}}
 ={\dot q}_{\k(\lambda)}.
\end{equation}
Therefore, the field Hamiltonian is
\begin{align}
 H &\equiv \int \d^3\k~\eta_{\lambda\lambda'}
    \biggl(p_{\k}^{(\lambda)} {\dot q}_{\k}^{(\lambda')}
    +p_{-\k}^{(\lambda)} {\dot q}_{-\k}^{(\lambda')}-L_{\k}\biggr) \\
  &= \int \d^3\k~\eta_{\lambda\lambda'}
 \biggl(p_{\k}^{(\lambda)} p_{-\k}^{(\lambda')}
 +|\k|^2 q_{\k}^{(\lambda)} q_{-\k}^{(\lambda')}\biggr) \label{eqn:Hamiltonian}.
\end{align}
Using this canonical setup of classical electromagnetism,
we eliminate the redundant variables due to the gauge constraint [\cref{Lorenz}],
and analyze the effects of incorporating this redundancy in the functional form
of the Poisson bracket corresponding to the field variables, $(A^{\mu},\pi^{\mu})$.
\section{Lorenz gauge and Poisson bracket}\label{sec:PB}
The Lorenz gauge condition in \cref{Lorenz}, when written in terms of the
dynamical variables,
$(q_{\pm\vb{k}}^{(\lambda)},p_{\pm\vb{k}}^{(\lambda)})$, becomes
\begin{equation}\label{eqn:GaugeCondition1}
 \l(\km_{i} \varepsilon^{i}_{(\lambda)}\r) q^{(\lambda)}_{\pm\k}
 \pm i\varepsilon^{0}_{(\lambda)} p^{(\lambda)}_{\mp\k}=0,
\end{equation}
which are a set of algebraic constraints between the variables
$q_{\pm\vb{k}}^{(\lambda)}$ and $p_{\pm\vb{k}}^{(\lambda)}$.

In the Hamiltonian view the gauge constraints in \cref{eqn:GaugeCondition1}
are imposed at one instant of time; it is then necessary to ensure that
they remain true under the time evolution generated by the Hamiltonian $H$.
That is, one needs to demand that
\begin{equation}\label{eqn:PBLorenz}
 \{\pa_{\mu}A^{\mu},H\}=0,
\end{equation}
with the expression for Poisson bracket (see \cref{app:CanT}) as
\begin{align}\label{eqn:PBDefinition}
 \{f_{\k},g_{\k'}\} &\equiv \eta^{\lambda\lambda'}
 \biggl(\f{\pa f_{\k}}{\pa q^{(\lambda)}_{\k}}
 \f{\pa g_{\k'}}{\pa p^{(\lambda')}_{\k}}
 -\f{\pa g_{\k'}}{\pa q^{(\lambda)}_{\k}}
 \f{\pa f_{\k}}{\pa p^{(\lambda')}_{\k}} \nn \\
 & \hskip -7pt
 + \f{\pa f_{\k}}{\pa q^{(\lambda)}_{-\k}}
 \f{\pa g_{\k'}}{\pa p^{(\lambda')}_{-\k}}
 -\f{\pa g_{\k'}}{\pa q^{(\lambda)}_{-\k}}
 \f{\pa f_{\k}}{\pa p^{(\lambda')}_{-\k}}\biggr) \delta(\k-\k'),
\end{align}
and it will generate an independent set of algebraic constraints in terms
of the dynamical variables,
$(q_{\pm\vb{k}}^{(\lambda)},p_{\pm\vb{k}}^{(\lambda)})$.

Using these expressions, and the field Hamiltonian, $H$,
[\cref{eqn:Hamiltonian}], to compute the Poisson bracket in 
\cref{eqn:PBLorenz} implies the algebraic constraints
\begin{equation}\label{eqn:GaugeCondition2}
 |\k|^2 \varepsilon^{0}_{(\lambda)}  q^{(\lambda)}_{\pm\k} 
 \pm i\l(\km_{i} \varepsilon^{i}_{(\lambda)}\r) p^{(\lambda)}_{\mp\k}=0.
\end{equation}
One can further check whether the condition imposed before, i.e.,
$\{\pa_{\mu}A^{\mu},H\}=0$, too remains enforced in time, which means demanding
$\{\{\pa_{\mu}A^{\mu},H\},H\} =0$. Direct computation yields
\begin{equation}\label{eqn:GaugeCondition3}
\begin{split}
 \{\{\pa_{\mu}A^{\mu},H\},H\} & = \int \d^3\k ~ e^{-i \vb{k}\cdot\vb{x}} \\
 & \times \l[\l(\km_{i} \varepsilon^{i}_{(\lambda)}\r) q^{(\lambda)}_{\pm\k}
 \pm i\varepsilon^{0}_{(\lambda)} p^{(\lambda)}_{\mp\k}\r].
\end{split}
\end{equation}
Comparing the expression in the square bracket with the algebraic equation
in \cref{eqn:GaugeCondition1}, it is evident 
that the condition, $\{\{\pa_{\mu}A^{\mu},H\},H\}=0$, does not need to be
demanded separately, since the algebraic equation
in \cref{eqn:GaugeCondition1} makes it inevitable.

Since the Lorenz gauge constraint [\cref{Lorenz}] is manifested as two
pairs of algebraic equations [\cref{eqn:GaugeCondition1,eqn:GaugeCondition2}],
eliminating the gauge constraint [\cref{Lorenz}] amounts to solving
these algebraic equations
to eliminate the redundant variables,
say, $(q^{(0)}_{\pm\k},p^{(0)}_{\pm\k})$. Therefore, the redundant variables,
$(q^{(0)}_{\pm\k},p^{(0)}_{\pm\k})$,
can be expressed in terms of the independent ones,
$(q^{(i)}_{\pm\k},p^{(i)}_{\pm\k})$, as
\begin{subequations}\label{eqn:q0p0}
\begin{align}
 q^{(0)}_{\pm\k} &= \f{\alpha_{(i')} q^{(i')}_{\pm\k}
 \mp i\beta_{(i')} p^{(i')}_{\mp\k}}{\gamma}, \\
 p^{(0)}_{\mp\k} &= \f{\pm i|\k|^2 \beta_{(i')} q^{(i')}_{\pm\k}
 + \alpha_{(i')} p^{(i')}_{\mp\k}}{\gamma},
\end{align}
\end{subequations}
where
\begin{subequations}\label{eqn:AlphaBetaGamma}
\begin{align}
 \alpha_{(i)} &\equiv -\l(\km_{\ell} \varepsilon^{\ell}_{(0)}\r)
 \l(\km_{m} \varepsilon^{m}_{(i)}\r)
 + |\k|^2 \varepsilon^{0}_{(0)} \varepsilon^{0}_{(i)}, \\
 \beta_{(i)} &\equiv \l(\km_{\ell}\varepsilon^{\ell}_{(0)}\r)
 \varepsilon^{0}_{(i)}
 - \varepsilon^{0}_{(0)} \l(\km_{m} \varepsilon^{m}_{(i)}\r), \\
 \gamma &\equiv \l(\km_{\ell} \varepsilon^{\ell}_{(0)}\r)^2
 - |\k|^2 \l(\varepsilon^{0}_{(0)}\r)^2.
\end{align}
\end{subequations}
Therefore, the gauge constraint in \cref{Lorenz} gets manifested
into four algebraic relations between the redundant variables,
$(q^{(0)}_{\pm\k},p^{(0)}_{\pm\k})$, and
the independent variables, $(q^{(i)}_{\pm\k},p^{(i)}_{\pm\k})$,
as shown in \cref{eqn:q0p0}.
These four algebraic relations [\cref{eqn:q0p0}] reduce the total
number of degrees of freedom,
$(q^{(\lambda)}_{\pm\k},p^{(\lambda)}_{\pm\k})$, from sixteen
(when gauge constraints are not imposed),
to twelve. Since the variables, $(q^{(i)}_{\pm\k},p^{(j)}_{\pm\k})$,
are supposed to combine in describing
the six components of
electric and magnetic fields, $({\bf E},{\bf B})$, it may appear
that the (twelve) independent variables,
$(q^{(i)}_{\pm\k},p^{(j)}_{\pm\k})$, over-describe the system, and
it can be shown straightforwardly that this
is not the case (see appendix \ref{app:VarCount}).

Therefore, the linearly independent canonical variables,
$(q^{(i)}_{\pm\k},p^{(i)}_{\pm\k})$,
that are left after taking the gauge constraint [\cref{Lorenz}]
into consideration, possess a
Poisson bracket expression in a reduced form as
\begin{align}\label{eqn:PBnew}
   \{f_{\k},g_{\k'}\}_{\mathbf{r}} &\equiv \delta^{ij}
   \biggl(\f{\pa f_{\k}}{\pa q^{(i)}_{\k}}
   \f{\pa g_{\k'}}{\pa p^{(j)}_{\k}}
   -\f{\pa g_{\k}'}{\pa q^{(i)}_{\k}} \f{\pa f_{\k}}{\pa p^{(j)}_{\k}} \nn \\
   &+ \f{\pa f_{\k}}{\pa q^{(i)}_{-\k}} \f{\pa g_{\k'}}{\pa p^{(j)}_{-\k}}
   -\f{\pa g_{\k}'}{\pa q^{(i)}_{-\k}}
   \f{\pa f_{\k}}{\pa p^{(j)}_{-\k}} \biggr)\delta(\k-\k'),
\end{align}
which implies
\begin{subequations}\label{eq:PBqipi}
\begin{align}
 \{q^{(i)}_{\k},p^{(j)}_{\k'}\}_{\mathbf{r}}
 &= \delta^{ij} \delta(\k-\k'), \\
 \{q^{(i)}_{\k},p^{(j)}_{-\k'}\}_{\mathbf{r}}
 &= \{q^{(i)}_{-\k},p^{(j)}_{\k'}\}_{\mathbf{r}}=0, \\
 \{q^{(i)}_{\pm\k},q^{(j)}_{\pm\k'}\}_{\mathbf{r}}
 &= \{p^{(i)}_{\pm\k},p^{(j)}_{\pm\k'}\}_{\mathbf{r}}=0,
\end{align}
\end{subequations}
where we call $\{\cdot,\cdot\}_{\mathbf{r}}$ as the reduced--Poisson
bracket corresponding to the canonical variables $(q^{(i)}_{\pm\k},p^{(i)}_{\pm\k})$.
This is also evident from \cref{eqn:PBDefinition}, since any general
observable, after taking the gauge constraint
[\cref{eqn:GaugeCondition1,eqn:GaugeCondition2}] into consideration, are functions
only of the dynamical variables $(q^{(i)}_{\pm\k},p^{(i)}_{\pm\k})$.

Since the (redundant) variables, $(q^{(0)}_{\pm\k},p^{(0)}_{\pm\k})$, 
depend on the variables, $(q^{(i)}_{\pm\k},p^{(i)}_{\pm\k})$, as given
by \cref{eqn:q0p0}, the reduced-Poisson bracket of the former can be evaluated to be
\begin{equation}\label{eq:PBq0p0}
 \{q^{(0)}_{-\k},p^{(0)}_{-\k'}\}_{\mathbf{r}}
 =\{q^{(0)}_{\k},p^{(0)}_{\k'}\}_{\mathbf{r}}
 = \l(\f{2|\k|^2}{\gamma}+1\r) \delta(\k-\k').
\end{equation}
The Poisson brackets of these dynamical variables [i.e. \cref{eq:PBqipi,eq:PBq0p0}]
can be used explicitly to compute the Poisson bracket of the field variable
$A^{\mu}$, and its conjugate momentum variable $\pi^{\mu}$ [\cref{eqn:pi}]:
\begin{align}
 \{A^{\mu}(t,\x),\pi^{\nu}(t,\x')\} &= \int \d^3\k
 ~\d^3\k'~\e^{-i\k \cdot \x}~\e^{i\k'\x'} \nn \\
 &\times \varepsilon^{\mu}_{(\lambda)} \varepsilon^{\nu}_{(\lambda')}
 \{q^{(\lambda)}_{\k},p^{(\lambda')}_{\k'}\}_{\mathbf{r}},
\end{align}
which is evaluated to be
\begin{align}\label{eqn:PBComplete}
 &\{A^{\mu}(t,\x),\pi^{\nu}(t,\x')\}
 = \int \d^3\k~\e^{-i\k \cdot (\x-\x')} \nn \\
 &\times \biggl\{\eta^{\mu\nu}
 + \f{1}{\gamma} \biggl[-k_{\ell} \l(\eta^{\ell\mu} \varepsilon^{\nu}_{(0)}
 +\eta^{\ell\nu} \varepsilon^{\mu}_{(0)}\r)
 \l(k_{m} \varepsilon^{m}_{(0)}\r) \nn \\
 &+ |\k|^2 \varepsilon^{0}_{(0)}
 \l(\varepsilon^{\mu}_{(0)} \eta^{\nu 0}+\varepsilon^{\nu}_{(0)} \eta^{\mu 0}\r)
 + 2|\k|^2 \varepsilon^{\mu}_{(0)} \varepsilon^{\nu}_{(0)}\biggr]\biggr\}.
\end{align}
\subsection{What preserves the Poisson bracket?}

Though the Lorenz gauge constraint [\cref{Lorenz}] apparently modifies the
Poisson bracket of the field and its conjugate momentum in a nontrivial manner
[\cref{eqn:PBComplete}], it turns out that the principle of relativity reduces
the Poisson bracket expression in \cref{eqn:PBComplete} to the simpler form as
in \cref{PB1}. According to the principle of relativity, there cannot exist a
rest frame for electromagnetic fields, since it is a massless (gauge) field;
implying that the polarization vector $\varepsilon^{\mu}_{(0)}$ is
\cite{Greiner,cohen}
\begin{equation}\label{eqn:lambda=0}
 \varepsilon^{\mu}_{(0)}=\l(-1,0,0,0\r).
\end{equation}
Remarkably, making use {\it only} of this fact, \cref{eqn:PBComplete}
reduces nicely to
\begin{equation}\label{eqn:PB-Structure}
 \{A^{\mu}(t,\x),\pi^{\nu}(t,\x')\}=\eta^{\mu\nu} \delta(\x-\x'). 
\end{equation}
Therefore, the principle of relativity plays an essential role in ensuring
the validity of the Poisson bracket relation in \cref{PB1}, even {\it after}
the Lorenz gauge constraint [\cref{Lorenz}] is imposed.
\section{Conclusion}\label{sec:Conclusion}
If the field variables, $A^{\mu}$, of classical electromagnetism were
dynamically independent, then the functional form of the Poisson bracket
between the field, $A^{\mu}$, and its conjugate momentum, $\pi^{\mu}$, would
have been of the form given in \cref{PB1}. However, since the field variables,
$A^{\mu}$, are related by a gauge constraint [\cref{Lorenz}], the Poisson
bracket is modified to the form in \cref{eqn:PBComplete}. But, remarkably,
the principle of special relativity --- which obviates a rest frame for
electromagnetic fields and consequently implies that the polarization vector
satisfies \cref{eqn:lambda=0} --- suffices to ensure that the Poisson bracket
in \cref{eqn:PBComplete} takes the oft-assumed form in \cref{PB1}. This state
of affairs is a result of the careful interplay between relativity, gauge
invariance, and the Hamiltonian structure of classical electromagnetism
--- an analysis of which seems to be missing in the extant literature,
to the best of our knowledge, and which we have provided in a pedagogical fashion.
\section{Acknowledgments}
I would like to thank Vivishek Sudhir for helpful comments,
and interesting discussions.
\appendix
\section{Canonical variables, $(q_{\pm\vb{k}}^{(\lambda)},p_{\pm\vb{k}}^{(\lambda)})$,
and their real and imaginary parts}\label{app:CanT}
The variables, $(q_{\pm\vb{k}}^{(\lambda)},p_{\pm\vb{k}}^{(\lambda)})$,
are complex, and they indeed posses a Poisson bracket definition as
\begin{align}
 \{f_{\k},g_{\k'}\} &\equiv \eta^{\lambda\lambda'}
 \biggl(\f{\pa f_{\k}}{\pa q^{(\lambda)}_{\k}}
 \f{\pa g_{\k'}}{\pa p^{(\lambda')}_{\k}}
 -\f{\pa g_{\k'}}{\pa q^{(\lambda)}_{\k}}
 \f{\pa f_{\k}}{\pa p^{(\lambda')}_{\k}} \nn \\
 & \hskip -7pt
 + \f{\pa f_{\k}}{\pa q^{(\lambda)}_{-\k}}
 \f{\pa g_{\k'}}{\pa p^{(\lambda')}_{-\k}}
 -\f{\pa g_{\k'}}{\pa q^{(\lambda)}_{-\k}}
 \f{\pa f_{\k}}{\pa p^{(\lambda')}_{-\k}}\biggr) \delta(\k-\k'),
\end{align}
which satisfies the Poisson bracket relations as
\begin{align}
 \{q^{(\lambda)}_{\k},p^{(\lambda')}_{\k'}\}
 &= \{q^{(\lambda)}_{-\k},p^{(\lambda')}_{-\k'}\}
 =\eta^{\lambda\lambda'} \delta(\k-\k'), \\
 \{q^{(\lambda)}_{\k},p^{(\lambda')}_{-\k'}\}
 &=\{q^{(\lambda)}_{-\k},p^{(\lambda')}_{\k'}\} =0, \\
 \{q^{(\lambda)}_{\pm\k},q^{(\lambda')}_{\pm\k'}\}
 &=\{p^{(\lambda)}_{\pm\k},p^{(\lambda')}_{\pm\k'}\} =0.
\end{align}
If the variables,
$(q_{\pm\vb{k}}^{(\lambda)},p_{\pm\vb{k}}^{(\lambda)})$,
are expressed in terms of its real and imaginary parts
in a particular combination as
\begin{equation}\label{eqn:CanT}
 q^{(\lambda)}_{\pm\k}=\f{{\tilde q}^{(\lambda)}_{\k}
 \pm iQ^{(\lambda)}_{\k}}{\sqrt{2}}, ~~~~
 p^{(\lambda)}_{\pm\k}=\f{{\tilde p}^{(\lambda)}_{\k}
 \mp iP^{(\lambda)}_{\k}}{\sqrt{2}},
\end{equation}
then the (real) variables,
$({\tilde q}^{(\lambda)}_{\k}, {\tilde P}^{(\lambda)}_{\k},
Q^{(\lambda)}_{\k},{\tilde P}^{(\lambda)}_{\k})$,
satisfy the Poisson bracket relation as
\begin{align}
 &\{{\tilde q}^{(\lambda)}_{\k},{\tilde p}^{(\lambda')}_{\k'}\}
 =\{Q^{(\lambda)}_{\k},P^{(\lambda')}_{\k'}\}
 =\eta^{\lambda\lambda'} \delta(\k-\k'), \\
 &\{{\tilde q}^{(\lambda)}_{\k},{\tilde q}^{(\lambda')}_{\k'}\}
 =\{{\tilde p}^{(\lambda)}_{\k},{\tilde p}^{(\lambda')}_{\k'}\} = 0, \\
 &\{Q^{(\lambda)}_{\k},Q^{(\lambda')}_{\k'}\}
 =\{P^{(\lambda)}_{\k},P^{(\lambda')}_{\k'}\}=0, \\
 &\{{\tilde q}^{(\lambda)}_{\k},Q^{(\lambda')}_{\k'}\}
 =\{{\tilde q}^{(\lambda)}_{\k},P^{(\lambda')}_{\k'}\}=0, \\
 &\{{\tilde p}^{(\lambda)}_{\k},Q^{(\lambda')}_{\k'}\}
 =\{{\tilde p}^{(\lambda)}_{\k},P^{(\lambda')}_{\k'}\}=0,
\end{align}
with the Poisson bracket definition in terms of the (real) variables
$({\tilde q}_{\vb{k}}^{(\lambda)},{\tilde p}_{\vb{k}}^{(\lambda)},
Q_{\vb{k}}^{(\lambda)},P_{\vb{k}}^{(\lambda)})$ as
\begin{align}\label{eqn:app:PBDef}
 &\{f_{\k},g_{\k'}\} \equiv \eta^{\lambda\lambda'}
 \biggl(\f{\pa f_{\k}}{\pa {\tilde q}^{(\lambda)}_{\k}}
 \f{\pa g_{\k'}}{\pa {\tilde p}^{(\lambda')}_{\k}}
 -\f{\pa g_{\k}'}{\pa {\tilde q}^{(\lambda)}_{\k}}
 \f{\pa f_{\k}}{\pa {\tilde p}^{(\lambda')}_{\k}} \nn \\
 &+ \f{\pa f_{\k}}{\pa Q^{(\lambda)}_{\k}}
 \f{\pa g_{\k'}}{\pa P^{(\lambda')}_{\k}}
 - \f{\pa g_{\k'}}{\pa Q^{(\lambda)}_{\k}}
 \f{\pa f_{\k}}{\pa P^{(\lambda')}_{\k}} \biggr)\delta(\k-\k'),
\end{align}
since the two sets of variables,
$(q_{\pm\vb{k}}^{(\lambda)},p_{\pm\vb{k}}^{(\lambda)})$
and $({\tilde q}_{\vb{k}}^{(\lambda)},{\tilde p}_{\vb{k}}^{(\lambda)},
Q_{\vb{k}}^{(\lambda)},P_{\vb{k}}^{(\lambda)})$, as related to each
other in \cref{eqn:CanT} is a canonical transformation.
\section{Counting the degrees of freedom}\label{app:VarCount}
Expressing the electric and magnetic fields, $(E^{i},B^{i})$, as
\begin{subequations}\label{eqn:EiBi}
\begin{align}
 &E^{i} = \int \d^{3} \k \l(E^{i}_{\k} \e^{-i\k \cdot \x}
 + E^{i}_{-\k} \e^{i\k \cdot \x}\r), \\
 &B^{i} = \int \d^{3} \k \l(B^{i}_{\k} \e^{-i\k \cdot \x}
 + B^{i}_{-\k} \e^{i\k \cdot \x}\r),
\end{align}
\end{subequations}
with the (twelve) independent field modes, $(E^{i}_{\pm\k},B^{i}_{\pm\k})$,
linearly related to the (twelve) independent canonical variables,
$(q^{(i)}_{\pm\k},p^{(i)}_{\pm\k})$, as
\begin{align}
\hskip -10pt
\begin{pmatrix}
 E^{i}_{\k} \\ \\
 E^{i}_{-\k} \\ \\
 B^{i}_{\k} \\ \\
 B^{i}_{-\k}
\end{pmatrix}
&=
\begin{pmatrix}
 i\Gamma^{i}_{(i')} & 0 & 0 & \Delta^{i}_{(i')} \\ \\
 0 & -i\Gamma^{i}_{(i')} & \Delta^{i}_{(i')} & 0 \\ \\
 i {\tilde \Gamma}^{i}_{(i')} & 0 & 0 & {\tilde \Delta}^{i}_{(i')} \\ \\
 0 & -i {\tilde \Gamma}^{i}_{(i')} & {\tilde \Delta}^{i}_{(i')} & 0
\end{pmatrix}
\begin{pmatrix}
 q^{(i')}_{\k} \\ \\
 q^{(i')}_{-\k} \\ \\
 p^{(i')}_{\k} \\ \\
 p^{(i')}_{-\k}
\end{pmatrix},
\end{align}
where
\begin{subequations}
\begin{align}
 \Gamma^{i}_{(i')} &\equiv
 \f{k^{i} {\cal E}^{0}_{(0)} \alpha_{(i')} - {\cal E}^{i}_{(0)} |\k|^2 \beta_{(i')}}{\gamma}
 - k^{i} {\cal E}^{0}_{(i')}, \\
 \Delta^{i}_{(i')} &\equiv
 \f{k^{i} {\cal E}^{0}_{(0)} \beta_{(i')} - {\cal E}^{i}_{(0)} \alpha_{(i')}}{\gamma}
 - {\cal E}^{i}_{(i')}, \\
 {\tilde \Gamma}^{i}_{(i')}
 &\equiv \f{\epsilon^{i}_{mn}k^{m}{\cal E}^{n}_{(0)}}{\gamma} \alpha_{(i')}
 + \epsilon^{i}_{mn} k^{m} {\cal E}^{n}_{(i')}, \\
 {\tilde \Delta}^{i}_{(i')} &\equiv
 \f{\epsilon^{i}_{mn}k^{m}{\cal E}^{n}_{(0)}}{\gamma} \beta_{(i')}.
\end{align}
\end{subequations}
Although the twelve linearly independent field modes,
$(E^{i}_{\pm\k},B^{i}_{\pm\k})$, in Eq.\ref{eqn:EiBi} are complex in
nature, they combine with the mode functions, $\e^{\pm i\k \cdot \x}$,
in such a way that the physical fields, $(E^{i},B^{i})$, are purely real.
In other words, the condition that the physical fields, $(E^{i},B^{i})$,
[\cref{eqn:EiBi}] to be real imposes a further restriction on the dynamical
variables, $(q^{(i)}_{\pm\k},p^{(i)}_{\pm\k})$, as
\begin{subequations}\label{eqn:qp_res}
\begin{align}
 q^{(i)}_{-\k} &= -\e^{-2i\k \cdot \x} q^{(i)}_{\k}, \\
 p^{(i)}_{-\k} &= \e^{2i\k \cdot \x} p^{(i)}_{\k}.
\end{align}
\end{subequations}
This again reduces the total number of independent variables from twelve
to six, which is indeed consistent with the number of independent field
components, $(E^{i},B^{i})$. Taking this restriction [\cref{eqn:qp_res}]
into consideration in \cref{eqn:EiBi}, the field components,
$(E^{i},B^{i})$, can be shown to be linearly related to the (six)
independent variables, $(q^{(i)}_{\k},p^{(i)}_{\k})$, as
\begin{align}
\hskip -9pt
 \begin{pmatrix}
  E^{i} \\ \\
  B^{i}
 \end{pmatrix}
 = - \hskip -5pt \int \d^3\k
 \begin{pmatrix}
  i\Gamma^{i}_{(i')} \e^{-i\k \cdot \x}
  & ~\Delta^{i}_{(i')} \e^{i\k \cdot \x} \\ \\
 i{\tilde \Gamma}^{i}_{(i')} \e^{-i\k \cdot \x}
 & ~{\tilde \Delta}^{i}_{(i')} \e^{i\k \cdot \x}
 \end{pmatrix}
 \hskip -5pt
 \begin{pmatrix}
  q^{(i')}_{\k} \\ \\
  p^{(i')}_{\k}
 \end{pmatrix}.
\end{align}


\begin{thebibliography}{50}
\bibitem{Whitt1917}
Whittaker,~E.~T., {\it A Treatise on the Analytical Dynamics of Particles and
Rigid Bodies} (Cambridge University Press, 1917).
%
\bibitem{Sommer1964}
Sommerfeld,~A., {\it Lectures on Theoretical Physics: Mechanics}
(Academic Press, 1964).
%
\bibitem{LanLif1976}
Landau,~L.~D. and Lifshitz,~E.~M., {\it Course of Theoretical Physics: Mechanics}
(Butterworth-Heinemann, 1976).
%
\bibitem{Arnold1989}
Arnold,~V.~I., {\it Mathematical methods of classical mechanics}
(Springer, 1989).
%
\bibitem{Dirac58}
Dirac,~P.~A.~M., Generalized Hamiltonian Dynamics,
\href{http://rspa.royalsocietypublishing.org/cgi/doi/10.1098/rspa.1958.0141}
{{\it Proceedings of the Royal Society A}, {\bf 246}, 326 (1958)}.
%
\bibitem{Dirac-1964}
Dirac,~P.~A.~M., {\it Lectures on Quantum Mechanics}
(Dover Publications, 1964).
%
\bibitem{Teitelboim:1976}
Hanson,~T.,~Regge~A., and Teitelboim,~C., {\it Constrained Hamiltonian Systems}
(Accademia Nazionale dei Lincei, 1976).
%
\bibitem{Sund1982}
Sundermeyer,~K., {\it Constrained Dynamics} (Springer, 1982).
%
\bibitem{Sudarsan-Mukunda:2015}
Sudarshan,~E.~C.~G. and Mukunda,~N.,{\it Classical Dynamics: A Modern Perspective}
(World Scientific Publishing, Reprint Edition, 2015).
%
\bibitem{Greiner}
Greiner,~W., {\it Field quantization} (Springer-Verlag, Berlin, Heidelberg, 1996).
%
\bibitem{Itzykson-Zuber:2006}
Itzykson,~C. and Zuber,~J., {\it Quantum Field Theory} (Dover Publications, 2006).
%
\bibitem{Prokhorov:1988}
Prokhorov,~L.~V., Quantization of the electromagnetic field,
\href{https://iopscience.iop.org/article/10.1070/PU1988v031n02ABEH005699/meta}
{{\it Soviet Physics Uspekhi}, {\bf 31}, 151 (1988)}.
%
\bibitem{Scherer:1987}
Barcelos-Neto,~A., Das,~J., and Scherer,~W., Canonical quantization of constrained systems,
{\it Acta Physica Polonica}, {\bf B18}, 269 (1987).
%
\bibitem{GitTyut}
Gitman,~D.~M. and Tyutin,~I.~V., {\it Quantization of Fields with Constraints}
(Springer-Verlag, 1990).
%
\bibitem{Gieres:2021}
Blaschke,~D.~N. and Gieres,~F., On the canonical formulation of gauge field theories
and Poincare transformations, \href{10.1016/j.nuclphysb.2021.115366}
{{\it Nuclear Physics B}, {\bf 965}, 115366 (2021)}.
%
\bibitem{cohen}
Cohen-Tannoudji,~C., Dupont-Roc,~J., and Grynberg,~G.,
{\it Photons and Atoms: Introduction to Quantum Electrodynamics}
(Wiley Professional Paperback Edition, 1997).
\end{thebibliography}
\end{document}